\documentclass{basi}
\usepackage{epsfig}
\begin{document}
\title[Study of grain compositions in comet Levy 1990XX]{Study of grain compositions in comet Levy 1990XX } 

\author[H. S. Das et al]
       {H. S. Das,$^1$\thanks{e-mail: hs\_das@rediffmail.com} 
       R. Bhattacharjee,$^2$  B. K. Sinha,$^3$
        A. K. Sen,$^4$ \\
        $^1$Department of Physics, Kokrajhar Govt. College, Kokrajhar 783370, Assam, India\\
        $^2$\\Department of Physics, R. K. Mahavidyalaya, Kailashahar
	799277, Tripura, India\\
        $^3$Department of Applied Mathematics, U. N. S. I. E. T., V. B. S. Purvanchal University, 
\\Jaunpur, Uttar Pradesh, India\\
	$^4$Department of Physics, Assam University, Silchar
              788011, India
       }

\maketitle
\label{firstpage}
\begin{abstract}
  In the present work, the non-spherical dust grain
   characteristics of comet Levy 1990XX  with
   different silicate grain compositions (both pyroxene and
   olivine) are studied
   using the T-matrix method. Considering   amorphous pyroxene grain (Mg$_x$Fe$_{1-x}$SiO$_3$)
   with  $x$ (= 0.4, 0.5, 0.6, 0.7, 0.8, 0.95, 1) and amorphous
   olivine grain (Mg$_{2y}$Fe$_{2-2y}$SiO$_4$) with $y$ (= 0.4, 0.5) (Dorschner
   et al. 1995), the observed polarization
   data of comet Levy1990XX are analyzed using the T-matrix code at $\lambda$ = 0.485 $\mu
   m$.  The best-fit spherical volume equivalent values of effective radius ($r_{eff}$),
    effective variance ($v_{eff}$) and shape parameter (E)  obtained for
    the observed polarization data of
    comet Levy 1990 XX at $\lambda = 0.485 \mu m$ for amorphous pyroxene
    grains  (Mg$_x$Fe$_{1-x}$SiO$_3$) with $x$ = 0.4, 0.5, 0.6, 0.7,
    0.8, 0.95 and 1 are (0.213, 0.0015, 0.486), (0.215, 0.0022,
    0.486), (0.215, 0.0019, 0.486), (0.217, 0.0034, 0.486), (0.219,
    0.0057, 0.486), (0.222, 0.0083, 0.486) and (0.230, 0.0092, 0.486)
    respectively. However, the amorphous olivine
   grains show a bad fit to the observed  polarization data.
   Considering crystalline olivine grain  in comet Levy 1990XX, Das \& Sen (2006)
   (our earlier work) already studied non-spherical  grain characteristics
   of that comet using the T-matrix method and found crystalline olivine
   prolate grains show satisfactory results. It is found that the shape
   parameter (E = 0.486) coming out from the present analysis for pyroxene grains
   is same with that of crystalline olivine grains. Thus the
   analysis of polarimetric data of comet Levy 1990XX at $\lambda$ = 0.485 $\mu
   m$ shows the presence of both olivine and pyroxene grains in that comet.

\end{abstract}


\section{Introduction}
\label{sec:intro}
Comets are the least processed and the pristine materials of early
solar nebula. Our knowledge of cometary dust comes from
polarimetric studies of comets, remote observation of IR spectral
features and the \textit{in situ} measurement of comets (e.g.,
comet Halley and the `Stardust mission'). The polarization
measurement of the scattered radiation gives valuable information
about the  shape, structure and sizes of the dust particles. Many
investigators (Bastien et al. 1986; Kikuchi et al. 1987, 1989;
Lamy et al. 1987, Le Borgne et al. 1987; Mukai et al. 1987; Sen et
al. 1991a, 1991b; Chernova et al. 1993; Joshi et al. 1997; Ganesh
et al. 1998; Manset \& Bastien 2000, Das et al. 2004 etc.) have
studied linear and circular polarization measurements of several
comets. These studies enriched  the knowledge about the dust grain
nature of comets.

Before the \textit{in situ} analysis of cometary dust was possible
in 1986, its composition was inferred from meteor spectra and
laboratory analysis of \textit{interplanetary dust particles}
(IDP) thought to be related to comets (Millman 1977, Rahe 1981).
But, the \textit{in situ} measurement of Halley, gave us the first
direct evidence of grain mass distribution (Mazets et al. 1987).
Lamy et al. (1987) analyzed the data for comet Halley from
spacecrafts VegaI, Vega II and Giotto. The important information
about the chemical composition of dust particles in comet Halley
has been obtained from the dust impact mass analyzer PUMA 1 and 2
on Vega and PIA on Giotto spacecrafts (Kissel et al. 1986a, 1986b;
Mazets et al. 1987; Lamy et al 1987). The \textit{in situ}
measurement of comet Halley indicated three classes of particles:
$(i)$ The lighter elements H, C, N and O indicative of organic
composition of grains  called `CHON' particles , $(ii)$
Carbonaceous chondrites of Type I (C I chondrites) and ($iii$) Mg,
Si and Fe, called \textit{silicates} (Clark et al. 1986).

The infrared measurement of comets has provided useful information
on the physical nature of cometary dust grains. Spectral features
at 10 $\mu m$ wavelength allowed the identification of silicates
in comet dust. Another silicate feature at 20 $\mu m$ also appears
to be present in many comets. The wavelength and shapes of these
feature provide important information for the identification of
the mineral composition (Wooden et al. 1997, 1999; Hanner 1999;
Hayward et al. 2000; Harker et al. 2002; Moreno et al. 2003 etc.).

The dust particles released by comets, are believed to contribute
to the population of \textit{interplanetary dust particles} (IDP),
often get collected at high altitudes of the Earth's atmosphere.
Laboratory studies have shown that majority of the collected IDPs
fall into one of the three spectral classes defined by their 10
$\mu m$ feature profiles. These observed profiles indicate the
presence of olivine, pyroxene and layer lattice silicates. This is
in agreement with the results obtained from Vega and Giotto mass
spectrometer observations of comet Halley (Lamy et al., 1987).
Mg-rich silicate crystals are also found within IDPs  and are
detected through cometary spectra (Hanner et al., 1999; Wooden et
al., 1999, 2000). If these Mg-rich crystals are solar nebula
remnants, then they must have formed in  an environment close to
the Sun (1 - 5 AU). Thus, the Mg-rich crystals represent grains
processed at high temperatures in the inner nebula. Therefore,
Mg-rich grains in comets probe physical conditions and processes
in the early solar nebula (Harker et al. 2002).

In the present work, the observed polarimetric data of comet Levy
1990XX at  $\lambda$ = 0.485 $\mu m$ (Chernova et al. 1993) have
been studied with different silicate grain compositions (both
pyroxene and olivine) using the T-matrix code.

\section{Spheroidal grain model and the T-matrix theory}
It is now accepted that cometary grains are not spherical and may be
\emph{fluffy aggregates} or \emph{porous}, with irregular or
spheroidal shapes (Greenberg $\&$ Hage 1990). The measurement of
circular polarization of comet Hale-Bopp (Rosenbush et al. 1997)
reveals that cometary grains must be composed of non-spherical
particles. In order to study the irregular grain characteristics of
comets, T-matrix theory (Waterman 1965)  is widely used by several
investigators.  Using the T-matrix code, Kerola $\&$ Larson (2001)
analyzed the polarization data of comet Hale-Bopp and found prolate
grains to be more satisfactory in that comet. Recently Das $\&$ Sen
(2006)(our earlier work), using the T-matrix code, discovered that
the prolate crystalline olivine grains can well explain the observed
polarization in a more satisfactory manner as compared to the other
shapes in comet Levy 1990XX.

The T-matrix method is a powerful exact technique used to study
the irregular grain characteristics of comet. In the present work,
calculation has been carried out for randomly oriented spheroids
using Mishchenko's (1998) single scattering  T-matrix code, which
is available at \textit{http://www.giss.nasa.gov/$\sim$ crmim}.
The main feature of the T-matrix approach is that it reduces
exactly to the Mie theory when the particle is a homogeneous or
layered sphere composed of isotropic materials. In the present
work, the \textit{power law size distribution} has been used for
the analysis of polarimetric data, where the minimum and maximum
particle radius are automatically set fixed for each and every run
merely by specifying the particle effective radius ($r_{eff}$) and
effective variance ($v_{eff}$).

The linear polarization of the scattered light by dust particles
depends upon $(i)$ wavelength of incident light ($\lambda$), $(ii)$
the geometrical shape (E) and size ($r$) of the particle, $(iii)$
Scattering angle, $\theta$ (= $180^0 -$ Phase angle),  and $(iv)$
the composition of dust particles in terms of complex values of
refractive index, $m$ $(= n - ik)$.  It is to be noted that  E $> 1$
for oblate spheroids, E $< 1$ for prolate spheroids and E $= 1$ for
spheres.

The dust particles released by comets, are believed to contribute
to the population of \textit{interplanetary dust particles} (IDP).
It has been observed that IDPs contain silicates both amorphous
and crystalline in structure (Bradley et al. 1999). The major
silicate species are \emph{pyroxene} (Mg$_x$Fe$_{1-x}$SiO$_3$,
where $0 \le x \le 1$) and \emph{olivine} (Mg$_{2y}$ Fe$_{2-2y}$
SiO$_4$, where $0 \le y \le 1$). A value of $x = y = 1$ denotes
the Mg-pure end member of the mineral and a value of $x = y = 0$
denotes the Fe-pure end member (Dorschner et al. 1995). It has
been observed that comet Levy 1990 XX displays a strong silicate
feature with a distinct peak at 11.25 $\mu m$, attributed to
crystalline olivine grains (Lynch et al. 1992).

In the present work, indices of refraction ($n,k)$ of amorphous
silicate minerals (pyroxene and olivine) (Dorschner et al. 1995)
have been taken. Since the refractive index parameters are not
available at 0.485$\mu m$, so the values are taken at 0.50 $\mu
m$. It is to be noted that a higher Mg-content corresponds to a
lower value of `$k$' (lower attenuation) in the UV, visible and
NIR; and the more Fe in the mineral, the stronger the absorptivity
in the UV, visible and NIR. Therefore, grains with less Fe and
more Mg will be poor absorbers of solar radiation and cooler than
their Fe-containing counter parts (Harker et al. 2002). Crystals
in \emph{Chondritic porous interplanetary dust particles} (CPIDPs)
(Bradley et al. 1996) are Mg-rich ($x\le 0.9$), solid and
submicron in size (Bradley et al. 1999), and of probable cometary
origin.

The polarimetric data of comet Levy 1990 XX has been taken from
Chernova et al. (1993). Since the polarimetric data is only
available at $\lambda = 0.485 \mu m$, our analysis is restricted
to that wavelength. Taking the grain composition of amorphous
pyroxene grain with $x=0.4$, the best fit values of
r$_{\mathrm{eff}}$, v$_{\mathrm{eff}}$ and E are determined using
the T-matrix code. The calculations are extended for  $x$ = 0.5,
0.6, 0.7, 0.8, 0.95 and 1 respectively. Now taking the grain
composition of amorphous olivine (with $y=0.4$ and $y=0.5$), the
calculation is repeated. The result coming out from the above
analysis is shown in \textbf{Table 1}. Also the result obtained
from our earlier work (Das \& Sen 2006) for crystalline olivine
grain is also shown.

It can be seen from \textbf{Table 1} that pyroxene glasses with
different grain compositions show significantly good fit to the
observed polarimetric data of comet Levy 1990XX. It is interesting
to note that the aspect ratio (E) coming out from the present analysis is same
for all the  grain compositions. Also, the effective radius of the pyroxene grains
increase with the increase of $x$. In \textbf{Fig 1}, the
simulated polarization curves with $x$ = 0.4, 0.5, 0.6, 0.7, 0.8,
0.95 and 1 are drawn  respectively on observed polarization data
of comet Levy 1990XX. The expected negative polarization curves
have been successfully generated for different values of $x$ using
the T-matrix code. It can be inferred that the analysis with amorphous olivine
grain shows bad fit to the observed data and is shown in
\textbf{Fig 2}. It is important to note that the expected negative polarization curves have not
been noticed with amorphous olivine grains. In \textbf{Fig 3}, the
polarization curves for amorphous pyroxene (with $x$ = 0.95 and 1)
and the curve obtained by Das \& Sen (2006) for crystalline
olivine grain are shown. It can be seen form \textbf{Table 1} that
crystalline olivine grains show a slightly better fit to the
observed polarimetric data as compared to amorphous pyroxene
grains.

\begin{figure*}
\centering
\includegraphics{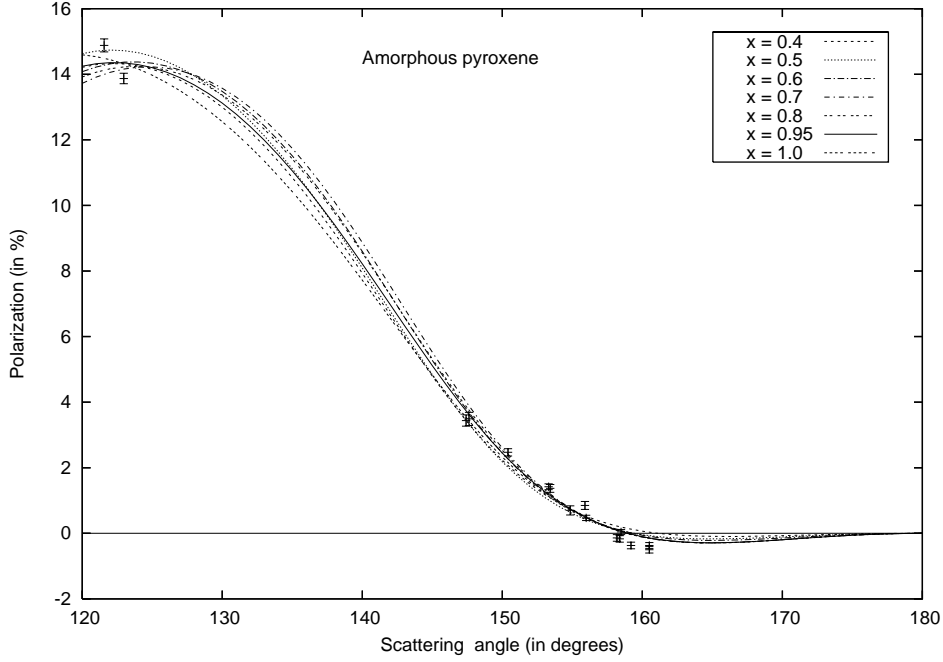}
\caption{The simulated
polarization curves generated using the T-matrix code at $\lambda$
= 0.485 $\mu m$ for amorphous pyroxene grain (Mg$_x$ Fe$_{1-x}$
SiO$_3$) with $x$ = 0.4, 0.5, 0.6, 0.7, 0.8, 0.95 and 1
respectively. }
\end{figure*}

\begin{figure*}
\centering
\includegraphics{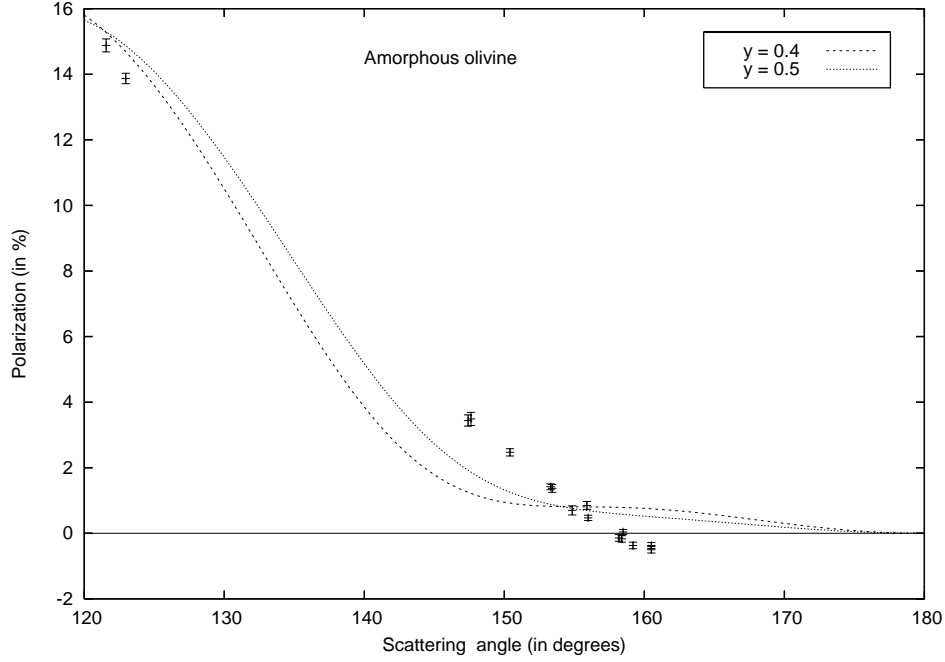}
\caption{The simulated
polarization curves generated using the T-matrix code at $\lambda$
= 0.485 $\mu m$ for amorphous olivine grain (Mg$_{2y}$ Fe$_{2-2y}$
SiO$_4$}
\end{figure*}

\begin{figure*}
\centering
\includegraphics{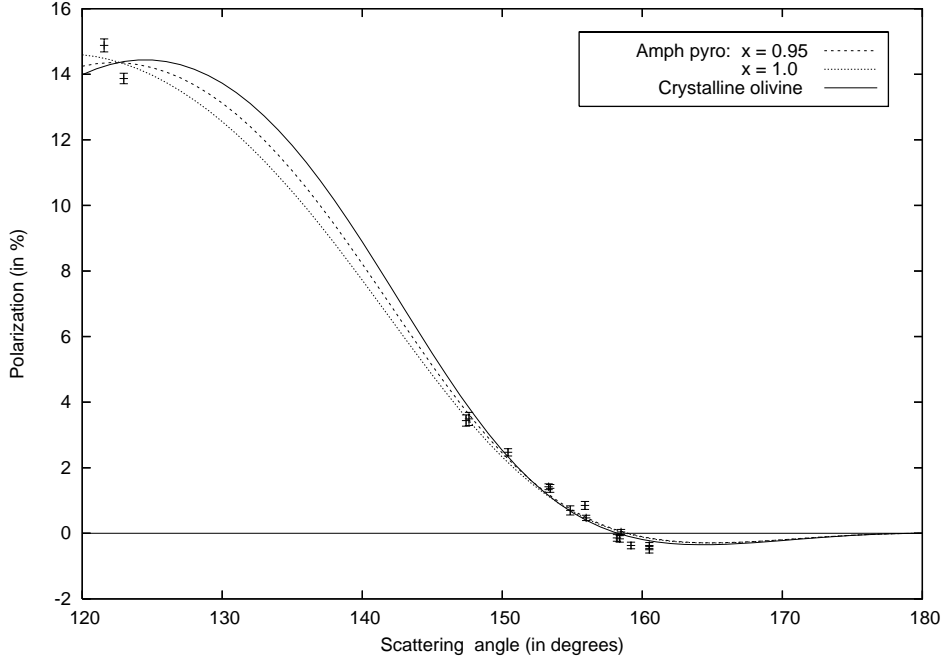}
\caption{The best fit
polarization curves for amorphous pyroxene with $x$ = 0.95 and 1
at $\lambda$ = 0.485 $\mu m$. The curve for crystalline olivine
grain from Das \& Sen (2006) is also drawn here. }
\end{figure*}
\begin{table}
\begin{center}
 \caption{The size distribution
   parameter (r$_{\mathrm{eff}}$, v$_{\mathrm{eff}})$ and shape parameter
   (E) at $\lambda = 0.485 \mu
   m$ for   pyroxene (amorphous)  and olivine (amorphous and
   crystalline) grains in comet Levy 1990XX. \label{tbl-1}}
   \hspace*{0.5cm}
\begin{tabular}{|c|l|l|l|c|c|c|c|c|}
  \hline
  Material&\multicolumn{2}{c}{Refractive index }\vline & $\, \, \;$x & y &E &
  $r_{\mathrm{eff}}$ & $v_{\mathrm{eff}}$ & $\chi ^2_{\mathrm{min}}$ \\
    \cline{2-3}
&  $\, \, \,$ $n$ & $\, \, \, \, \,$ $k$ & & & &($\mu m$) &  &\\
\hline
 \hline
 & 1.709 \hfill \hfill \hfill \hfill & 0.0515& 0.4 & - &0.486 & 0.213 & 0.0015 & 9.93\\
 \cline{2-9}
 &1.688&0.0448&0.5& -&0.486&0.215&0.0022&9.28\\
 \cline{2-9}
Amorphous Pyroxene&1.678&0.0291&0.6&- &0.486&0.215&0.0019&6.54\\
 \cline{2-9}
(Mg$_x$Fe$_{1-x}$SiO$_3$) &1.639&0.0057&0.7& -& 0.486&0.217&0.0034&5.84\\
 \cline{2-9}
 &1.613&0.00179&0.8&-&0.486&0.219&0.0057&5.46\\
 \cline{2-9}
 &1.589&0.00029&0.95&-&0.486&0.222&0.0083&5.30\\
 \cline{2-9}
 &1.577&0.00002&1&-&0.486&0.230&0.0092&5.35\\
 \cline{2-9}
 \hline
 \hline
 Amorphous Olivine & 1.823&0.104 &-&0.4 & 0.461&0.210 &0.0004 & 82.30 \\
 \cline{2-9}
 (Mg$_{2y}$Fe$_{2-2y}$SiO$_4$) & 1.768&0.101 &-& 0.5&0.474 &0.213 &0.0019 &50.69 \\
 \hline
 \hline
Crystalline Olivine & 1.63 & 0.00003& - & - &0.486 & 0.218 & 0.0036 & 5.22\\
(Source: Das \& Sen 2006)&&&&&&&&\\
\hline
\end{tabular}
\end{center}
\end{table}

\section{Discussions}
The study of the cometary spectra in the thermal region (8
- 45 $\mu m$) has revealed the presence of amorphous carbon and
amorphous silicates, as well as crystalline olivine (Campins \&
Ryan 1989, Lynch et al. 1992, Brucato et al. 1999, Wooden et al.
1999, Harker et al. 2002 etc.). These findings are consistent with
the composition of cometary dust particles as detected by Giotto
and Vega spacecrafts (Kissel et al. 1986a). The silicate emission
features are also noticed in the observations of comet Hale-Bopp
from ISO (Infrared Space Observatory) (Crovisier et al. 2000).
Mg-rich silicate crystals are also found within IDPs  and are
detected through cometary spectra (Hanner et al. 1999; Wooden et
al. 1999, 2000). Thus it is clear that Mg-rich grains are more
satisfactory in comets.

In the present study, the non-spherical grain characteristics of
comet Levy 1990XX are studied using the T-matrix code, considering
different silicate grain (amorphous and crystalline both)
compositions. It can be seen from \textbf{Table - 1} that Mg-rich
amorphous pyroxene grains show good fit to the observed polarimetric data 
compared to amorphous olivine grains. Das \& Sen (2006) already found that the crystalline olivine
grains also show a good fit to the observed data. Thus it can be
inferred from the present analysis that the dust grains in comet
Levy 1990 XX are mostly amorphous pyroxene and crystalline
olivine. It is interesting to note that the shape parameter coming
out from the present analysis is same for both pyroxene and
crystalline olivine grains and is given by 0.486. It has also been observed  that the
effective radii of the pyroxene  grains increase with the increase
of $x$ (Ref: \textbf{Table - 1}).

\section{Summary}
\begin{enumerate}
  \item The best-fit spherical volume equivalent values of $r_{eff}$,
    $v_{eff}$ and E  obtained for the observed polarization data of
    comet Levy 1990 XX, at $\lambda = 0.485 \mu m$ for amorphous pyroxene
    grains  (Mg$_x$Fe$_{1-x}$SiO$_3$) with $x$ = 0.4, 0.5, 0.6, 0.7,
    0.8, 0.95 and 1 are (0.213, 0.0015, 0.486), (0.215, 0.0022,
    0.486), (0.215, 0.0019, 0.486), (0.217, 0.0034, 0.486), (0.219,
    0.0057, 0.486), (0.222, 0.0083, 0.486) and (0.230, 0.0092, 0.486)
    respectively.

  \item  The expected negative polarization values have been successfully
         generated for prolate pyroxene grains using the T-matrix code.

  \item  The amorphous olivine grains show a bad fit to the
         observed data of comet Levy 1990 XX and the simulated polarization curves do not show 
	 any negative polarisation.
  \item  It can be inferred from the present analysis that the dust grains in comet
         Levy 1990 XX are mostly amorphous pyroxene and
         crystalline olivine with same aspect ratio (E = 0.486).
\end{enumerate}

\section*{Acknowledgements}

 The authors sincerely acknowledge IUCAA, Pune, where some part of
 these calculations were done. The authors are also thankful to M.
 Mishchenko for the T-matrix code.

\label{lastpage}
\end{document}